\documentclass[sigconf,nonacm]{acmart}

\usepackage[utf8]{inputenc}
\usepackage{amsmath}
\usepackage{amsthm}
\usepackage{xspace}

\newcommand{\mtoken}[1]{\ensuremath{\textsf{#1}}\xspace}

\newcommand{\original}{\ensuremath{x}\xspace}
\newcommand{\derived}{\ensuremath{y}\xspace}
\newcommand{\program}{\ensuremath{M}\xspace}
\newcommand{\producer}{\ensuremath{P}\xspace}
\newcommand{\reproducer}{\ensuremath{R}\xspace}
\newcommand{\plaintiff}{\mtoken{plaintiff}}
\newcommand{\defendant}{\mtoken{defendant}}
\newcommand{\simProd}{\ensuremath{w}\xspace}
\newcommand{\simRep}{\ensuremath{z}\xspace}
\newcommand{\simi}{\ensuremath{\cong}\xspace}
\newcommand{\copyr}{\mtoken{copy}}
\newcommand{\noncopy}{\mtoken{noncopy}}
\newcommand{\context}{\mtoken{context}}
\newcommand{\dl}[3]{\ensuremath{\mathcal{C}(#1,#2 \,|\, #3)}\xspace}
\newcommand{\simdl}[3]{\ensuremath{\tilde{\mathcal{C}}(#1,#2 \,|\, #3)}\xspace}
\newcommand{\LK}{\mtoken{LK}}
\newcommand{\levin}[2]{\ensuremath{\LK(#1 \,|\, #2)}\xspace}
\newcommand{\Der}{\mtoken{DerSim}}
\newcommand{\copyclose}[3]{\Der(#1, #2\,|\, #3)\xspace}
\newcommand{\DerSim}[3]{\copyclose{#1}{#2}{#3}}
\let\DerAdv=\DerSim
\newcommand{\DerEmp}{\mtoken{DerSimEmp}}
\newcommand{\DerSimEmp}[5]{\DerEmp(#1, #2, #3, #4\,|\,#5)}

\newcommand{\inp}{\ensuremath{b}\xspace}
\newcommand{\bg}{\mtoken{bg}}

\usepackage{amsmath,amsthm}
\newtheorem{theorem}{Theorem}[section]

\theoremstyle{definition}
\newtheorem{definition}[theorem]{Definition}

\let\gets=\leftarrow

\newcommand{\tabtitle}[1]{\textsc{\underline{#1}}:}

\begin{document}

\title[Formalizing Human Ingenuity]{Formalizing Human Ingenuity: A Quantitative Framework for Copyright Law's Substantial Similarity}
\author{Sarah Scheffler}
\email{sscheff@princeton.edu}
\affiliation{%
  \institution{Princeton University}
  \city{Princeton}
  \state{NJ}
  \country{USA}
}

\author{Eran Tromer}
\email{tromer@cs.columbia.edu}
\affiliation{%
  \institution{Columbia University}
  \city{New York}
  \state{NY}
  \country{USA}
}

\author{Mayank Varia}
\email{varia@bu.edu}
\affiliation{%
  \institution{Boston University}
  \city{Boston}
  \state{MA}
  \country{USA}
  \postcode{02140}
}

\begin{abstract}
A central notion in U.S.~copyright law is judging the \emph{substantial similarity} between an original and an (allegedly) derived work.
Capturing this notion has proven elusive, and the many approaches offered by case law and legal scholarship are often ill-defined, contradictory, or internally-inconsistent.

This work suggests that key parts of the substantial-similarity puzzle are amendable to modeling inspired by
theoretical computer science. Our proposed framework quantitatively evaluates how much ``novelty'' is needed to produce the derived work with access to the original work, versus reproducing it without access to the copyrighted elements of the original work. ``Novelty'' is captured by a computational notion of description length, in the spirit of Kolmogorov-Levin complexity, which is robust to mechanical transformations and availability of contextual information.

This results in an actionable framework that could be used by courts as an aid for deciding substantial similarity. We evaluate it on several pivotal cases in copyright law and observe that the results are consistent with the rulings, and are philosophically aligned with the abstraction-filtration-comparison test of \emph{Altai}.

\end{abstract}

\maketitle

\section{Introduction}

In 2021, the U.S. Supreme Court ruled on \emph{Google LLC v. Oracle America, Inc.} \cite{googacle}, a case involving copyright of computer code.
This ruling was the culmination of nearly a decade of litigation, and it exposed crucial ambiguities in interpretations of many fundamental aspects of copyright law surrounding 
the copyrightability of Application Programming Interfaces, 
the merger doctrine, and the dichotomy between ideas and expression in copyright law.
While this high-profile case was ultimately decided based on the fair use doctrine, the fundamental ambiguities impact a growing number of copyright and computer software cases.

U.S.\ copyright law is complex, and accordingly any specific copyright case can turn on any of a myriad of considerations.
Some copyright cases are straightforward matters of testimony or feasibility.
Some cases depend on whether the defendant had access to the plaintiff's intellectual work when producing its
But many cases turn on a ``test'' that compares the original and derived intellectual works to see if they are \emph{substantially similar} in their creative \emph{expression}.

These tests generally rely on two related concepts.
First, they emphasize
human expression in intellectual works (of literature, music, art, and other tangible media) as distinct from a fact, abstract concept, or scientific idea; only the \emph{expression} is copyrightable
\cite{baker_v_selden}.
Second, U.S.\ courts apply the standard of \emph{substantial similarity} to decide whether infringement has occurred;
at a high level, both literal copying or producing a derived work that depends too heavily on the original run afoul of copyright law.
These two considerations are intervowen: Judge Learned Hand famously depicted the challenge as determining
``when an imitator has gone beyond copying the `idea,' and has borrowed its `expression''' \cite{peter_pan}.

Alas, legal doctrine and case law are notoriously unclear on the definition of such substantial similarity.
Judges and legal scholars have lamented about the lack of a consistent test.
For example, Balganesh remarks that the ```substantial similarity' requirement['s] \ldots structure, scope, and purpose continue to confound courts and scholars,''
and that in spite of its ``centrality \ldots to copyright law, its complexity renders it a virtual black hole in copyright jurisprudence''
\cite{balganesh}.
Lim makes a similar comment, calling substantial similarity ``copyright infringement's black box'' \cite{lim2021saving}.
Roodhuyzen observes that 
``[t]here is a great amount of confusion among courts and commentators as to what the proper test is for determining whether two works are `substantially similar' so as to constitute copyright infringement'' \cite{roodhuyzen}.
And this is true in spite of the fact that there is no shortage of proposed tests by courts and scholars to systematize the substantial similarity requirement (which we will describe in \S \ref{sec:background}).
Indeed, Samuelson states that ``it is problematic that there are so many different tests and
so little guidance about which test to use when'' \cite{samuelson}
and Stanfield critiques that the tests
``are not a means to determine similarity, but rather a means to explain a finding of similarity that is determined in such a way that defies clear explanation''~ \cite{stanfield}.

\paragraph{Our objective}
The goal of this work is to create, justify, and use a new computer science-inspired framework that can test whether one work is substantially similar to another.
We emphasize upfront that having a test of substantial similarity does \emph{not} eliminate the complexity of copyright cases, and it is neither our claim nor our goal to ``automate'' the legal system.
On the contrary, our intention is to form a more objective, predictable standard of determining substantial similarity precisely in order to free up the courts' time to ponder more difficult questions, such as the access and fair use questions at the heart of \emph{Google v.\ Oracle} \cite{googacle}.

\paragraph{From law to computer science and back again}
So why the need for yet another test of substantial similarity?
We offer three responses to this question, two from a legal perspective and one from a computer science viewpoint.

From a legal perspective, we claim that previous works tried to solve too onerous of a problem: constructing a test that judges or juries
can use on their own to evaluate substantial similarity.
Instead, we suggest a framework that leverages the adversarial system of justice: the litigants should put forward their best argument as to why copyright infringement did/didn't occur.
We require that an impartial party can easily determine which argument carries the day, but we do not require that the plaintiff and defendant's tasks are simple.
The adversarial approach incentivizes both parties to present the best arguments they know,
and applies a quantitative metric to weigh these arguments.

The question then arises: what quantitative metric is appropriate?
Some natural candidates turn out unsuitable.
Notions of conditional entropy and mutual information reason about random variables, rather than individual strings, and thus require modeling a counterfactual probability distribution --- with little connection to copyright law.
Common string difference notions, such as edit distance and earth mover's distance, compare strings at too superficial a level, and are trivially foiled by reshuffling, translating to a different language, etc.
Instead, we utilize a \emph{minimum description length notion} (specifically, conditional Levin-Kolmogorov complexity \cite{levin1973universal,DBLP:journals/iandc/Levin84,kolmogorov1963tables,DBLP:books/daglib/0071317}), which is both string-based and, by definition, highly robust to mechanical transformations and external context.
(This metric is also, in principle, infeasible to compute precisely; however in the law it is perfectly acceptable for it to be challenging to craft a strong argument, and furthermore we argue that a ``fair'' upper bound is provided by the truthful history of producing the works.)

This brings us back to the legal perspective, to ascertain that the mathematical formalism is not merely elegant, but also adheres to the letter and spirit of the law.
Indeed, we show that our test agrees with several landmark U.S.\ Supreme Court decisions in copyright law.
Additionally, we discuss why our definition comports with concepts in U.S.\ copyright law such as the  idea-expression doctrine, avoids the overturned ``sweat of the brow'' doctrine, and more.

\paragraph{Overview of our description-length framework}

Our novel methodological framework reasons about substantial similarity (i.e., the amount of expression that was copied) by using description length.
Specifically, given an original work $\original$ and an allegedly-copied work $\derived$,
we require that the plaintiff and defendant present computer programs that generate $\derived$ with and without access to $\original$, respectively.
Intuitively, the plaintiff's program $\producer$ describes how she thinks the defendant infringed on her copyright by producing $\derived$ from $\original$, while the defendant's program $\reproducer$ describes how she constructed $\derived$ without access to $\original$.
Then, we define the \emph{derivation similarity} as the difference in description lengths of these two programs $\producer$ and $\reproducer$.
This metric captures
the advantage gained by utilizing the copyrightable elements of $\original$ in producing $\derived$.
By using a Levin-style description length \cite{DBLP:journals/iandc/Levin84,DBLP:books/daglib/0071317}, our framework accounts for both the program's description length and any brute-force searches performed.

The resulting definition (see Def.\ \ref{def:empirical}) has its roots in the abstraction-filtration-comparison method of \emph{Computer Associates, Inc. v. Altai} \cite{ca_v_altai}.  It is
is easy for an impartial judge to evaluate, and therefore we believe it can be used by the courts.
In fact, our framework \emph{requires} that its inputs be created by the existing court system.
While our framework reduces the large, nebulous question of ``what is derivative work?'' into a collection of smaller, better-defined questions, they cannot all be handled using our computer-science metric.
For instance, care 
and legal knowledge are
required to identify the inputs to our system, namely, the copyrightable and non-copyrightable aspects of $\original$ and $\derived$; we require the courts to adjudicate this task (see \S \ref{sub:copyright-def} for details).
This is consistent with the approach in \emph{filtering} where the court (not a jury) performs filtration \cite[\S 8.6.2]{osterberg} and often relies on expert testimony when doing so \cite{gates_v_bando}.
Once these inputs have been identified, our framework greatly simplifies the next step of \emph{comparison} once the filtration step has been completed.

\paragraph{Our contributions}

In summary, we make the following contributions in this work.
\begin{itemize}
\item In \S \ref{sec:background}, we provide a brief 
taxonomy of existing copyright infringement tests (after providing a primer on copyright law for computer scientists).

\item In \S \ref{sec:definition}, we contribute our quantitative metric of \emph{derivation similarity}, which quantifies the extent to which one work is substantially similar to another.

\item In \S \ref{sec:precedent}, we validate and justify our frameworks by showing that its reasoning is consistent with the decisions in a wide variety of landmark court cases involving copyright law over the past century.
Additional, hypothetical, stress-tests are discussed in Section~\ref{sec:synthetic}.

\item In \S \ref{sec:zk}, we provide a simple cryptographic algorithm that allows the plaintiff and defendant to compute our derivation similarity metric privately, so that only the result is revealed in the public record.
\end{itemize}

\paragraph{Limitations}
This paper only considers copyright law in the United States; though measuring the degree of similarity between two works is a core principle in any copyright law, so the ideas might apply in additional jurisdictions.
Our framework focuses on substantial similarity, and does not cover other aspects of copyright law such as the fair use doctrine or the Digital Millenium Copyright Act (DMCA) anti-circumvention measures.
Also, our framework presently does not encompass randomized algorithms, and only partially handles hidden information or trade secrets \cite{verification_dilemmas}; see \S \ref{sub:discussion} and \S \ref{sec:zk} for details.
For these and other reasons, this quantitative framework should not be used as a method for determining whether there is copyright infringement overall; rather, it answers only the narrow question of whether two works are substantially similar.

\paragraph{Relationship to prior works}
This works joins recent scholarship on using concepts and modeling principles from computer science to reason about law and policy questions in a new light \cite{USENIX:SchVar21,EC:GarGolVas20,FORC:CohenDMS21,PNAS:CohenN20,arXiv:JudFei22,PODS:Nissim21,Nissim2017bridging,Altman2021hybrid}.
Other classes of CS and Law cross-disciplinary research include developing and deploying new computer science tools for realizing  social or legal goals (e.g., \cite{USENIX:FPSGW18,SP:KMPQ21,verification_dilemmas,WPES:GoldwasserP17,MOBISYS:KohNB21}),
and policy discussions about the appropriate uses and limits of computer science technology in society (e.g., \cite{arXiv:KKLNQTV21,USENIX:KulMay21,SPW:FeigenbaumW18,Bellovin2014lawful,AbelsonABBBDGG015}).

Additionally, this work connects to a diverse landscape of court decisions and legal scholarship surrounding copyright law, which is too broad to do justice to in this space.
The United States Congress' power to grant copyright protections is listed in Article I, Section 8 of the U.S. Constitution,
with laws including the Copyright Act of 1976 \cite{copyright76}
leading to its current statutory form that is codified within
Title 17 of the U.S.\ Code \cite{title17}.
There are hundreds of Supreme Court and appellate court decisions that have shaped the way that courts interpret copyright law
(e.g., \cite{googacle,krofft_v_mcdonalds,arnstein_v_porter,roth_v_united,sheldon_v_metro,feist_v_rural,rosenthal_v_kalpakian,nichols_v_universal,peter_pan,ca_v_altai,crowe1992,southco_v_kanebridge,baker_v_selden,boisson_v_banian,bender_v_west,morrissey_v_pg,rosenthal_v_kalpakian,eckes,financial_v_moodys,kregos_v_ap,lotus_v_borland,sega_v_accolade});
we describe several landmark cases in \S \ref{sec:precedent} and show that our framework agrees with the decisions in all of these cases.
Additionally, there are several casebooks
(e.g., \cite{ipbook,field,dukecaselaw})
and law journal articles
(e.g., \cite{laroche,lemley,BalganeshMW,balganesh,latman,samuelson,verification_dilemmas,ParchomovskyG,roodhuyzen,lim})
that describe and reflect the modern understanding of copyright law.
In the next section, we explore the landscape of copyright law and the tests that have been previously proposed to make sense of the ``substantial similarity'' requirement.

\section{Background on Copyright Law}
\label{sec:background}

Title 17 of the U.S.~Code \S 102 
states that ``[c]opyright protection subsists \ldots in original works of authorship \ldots In no case does copyright protection for an original work of authorship extend to any idea, procedure, process, system, method of operation, concept, principle, or discovery, regardless of the form in which it is described, explained, illustrated, or embodied in such work'' \cite{17usc102}.
Originality is a requirement --- the original work must ``possess some creative spark'' \cite[l. 345]{feist_v_rural}, but ``the requisite level of creativity is extremely low; even a slight amount will suffice'' \cite[l. 345]{feist_v_rural}.
The copyright-holder holds the exclusive right to copy, distribute, or produce derivative works based on the copyrighted work \cite{17usc106}, with some exceptions described in 17 U.S.C. \S 107-122 (including ``fair use,'' which we do not discuss further here).
Thus, the usual topic in a copyright case is to determine whether or not the \emph{expression} in the defendant's work $\derived$ was ``derived from'' the expression within the plaintiff's copyrighted work $\original$ (whereas reuse of \emph{ideas} represented by those expressions is always permissible \cite{17usc102,feist_v_rural}).

Sometimes this question is straightforward.  Most instances when $\derived$ is a literal copy of $\original$ are found to be copyright infringement (although if two authors truly did create the same work independently, there was no infringement \cite{sheldon_v_metro,feist_v_rural,nichols_v_universal}).  However, most of the time $\derived$ is not a literal copy of $\original$.  Many courts require that there be \emph{substantial similarity} between the works to demonstrate copying, although there is no general agreement on what exactly this means (see e.g. \cite{cohen1986masking,lemley,nimmer1988structured,osterberg,fleming1969substantial,samuelson}).
Each court applies the idea slightly differently, and the approaches to these methods have changed over time.  We proceed to describe some of these tests for determining whether two works' expressions are substantially similar.

\subsection{Copyright tests}
\label{copyright-tests}

We provide only a brief summary of the most important existing copyright tests here; for more details see \cite{osterberg,samuelson,nimmer1988structured}.

\paragraph{The Ordinary Observer Test}
One of the most widely-applied copyright tests, the Second Circuit's approach in \emph{Arnstein v. Porter} \cite{arnstein_v_porter} in 1946 has been adopted with minor variations by the First, Third, Fifth, and Seventh circuits as well \cite[\S3.1]{osterberg}. Its analysis combines two elements:  The court must determine whether (1) the defendant copied from the plaintiff's copyrighted work, and (2) if that copying went so far as to be improper \cite[l. 468]{arnstein_v_porter}.  In determining the first question, Judge Frank wrote in the case that extreme similarities between the works were suspicious, even if evidence of access to the original work was slight: ``a case could occur in which the similarities were so striking that we would reverse a finding of no access, despite weak evidence of access (or no evidence thereof other than the similarities); and similarly as a to a finding of no illicit appropriation'' \cite[l. 469]{arnstein_v_porter}, and the latter question involves an ``ordinary observer'' to determine if the copying ``resulted in substantial similarity'' \cite[l. 608]{concrete_v_classic}.  
Samuelson describes difficulty in applying this test because the key terms are used differently in the two parts of the test \cite{samuelson}, and Lemley asserts that the the methods for resolving the two questions should be reversed: an ordinary observer could detect copying, but the question of whether or not that was improper could require expert testimony or analysis \cite{lemley}.

\paragraph{The Extrinsic/Intrinsic Test}
\emph{Sid \& Marty Krofft Television Productions, Inc. v. McDonald's Corp}, a 1977 case from the Ninth Circuit, separated out the ``extrinsic'' elements like the work's type, materials, subject matter, setting, etc, from ``the observations and impressions of the average reasonable reader and spectator'' \cite{krofft_v_mcdonalds,fox_v_stonesifer}, which they dub as the ``intrinsic'' portion of the test.  While originally intending to compare similarities in ideas and expression respectively, those tests were later interpreted to refer to ``objective'' and ``subjective'' similarities (see e.g. \cite{shaw_v_lindheim}).  The intrinsic or subjective part of the test is also highly similar to the ``total look and feel'' test from the same circuit in 1970, which essentially asks whether ``the work is recognizable by an ordinary observer as having been taken from the copyrighted source'' \cite[l. 1110]{roth_v_united}.  These approaches have been criticized for making it too easy to include uncopyrightable elements in the similarity analysis \cite{samuelson} and that the overall ``feel'' of a work may make sense in the case of a small work like a greeting card, but not in a larger work like a large body of computer code \cite{nimmer1988structured}.
Nevertheless, the Fourth and Eighth Circuits both perform some variant of this test, and the Eleventh has done so for some cases \cite{osterberg}.

\paragraph{The Abstraction, Filtration, Comparison Test}
This test emerged from \emph{Computer Associates, Inc. v. Altai} for the purpose of determining the similarity of computer code.
In the framework, one begins by ``dissect[ing] the allegedly copied program's structure and isolat[ing] each level of abstraction contained within it,'' ending with ``an articulation of the program's ultimate function'' \cite[l. 707]{ca_v_altai} (this abstraction step had previously been applied to other works like plays \cite{nichols_v_universal}).  Secondly, one should conduct a ``successive filtering method'' \cite{ca_v_altai} to determine whether the inclusion of a particular element of one level of abstraction was due to ``idea'' or ``expression,'' and remove all the elements that were due to ideas, efficiency constraints, external compatibility, or from the public domain \cite[l. 709-710]{ca_v_altai}.  Finally, one should compare what is left after the filtration process is done --- all remaining similarities should be similiarities of expression.

Although this test was not without its critiques and wrinkles (e.g. \cite{butler1992pragmatism,effross1993assaying,heer2004case,    samuelson1994,softei_v_dragon,samuelson2016}),
the general approach has stood the test of time and is the primary tool for the courts to rule on software copyright cases \cite[\S8.6]{osterberg}.
It is used for all cases in the Tenth Circuit, and in the Sixth Circuit as well with a slight variation \cite[\S3.3]{osterberg}.  The D.C.\ Circuit also practices a filtration step before reverting to something more akin to the ordinary observer test \cite[\S3.3.3]{osterberg}.
Other courts have also adopted filtering tests like \emph{Altai} into other settings like advertisements \cite{transwestern_v_multimedia,ready_v_cantrell},  architecture \cite{architecture}, child safety locks \cite{kohus_v_mariol}, and dolls \cite{country_v_sheen}.

\subsection{Where our framework fits in}
\label{ssec:where-we-fit-in}

In the abstraction-filtration-comparison approach of \emph{Altai}, determining which elements may be ``filtered out'' is a challenging question requiring expertise in copyright law and the subject matter.
However even if the ``filtering'' step was done perfectly, it is still challenging to rigorously state in the ``comparison'' step how much illegal copying occurred.  This is often considered a ``value judgement, involving an assessment of the importance of the material that was copied'' \cite[Vol. 3 \S13.03]{nimmer}.  The copied expression must be ``important'' to the copied work even though it may not be a large portion of it \cite[\S8.6.3]{osterberg}.

Our approach provides a way to reduce the qualitative question of ``how important'' the copying was to a quantitative assessment that may be determined after the filtration step was completed.  We do \emph{not} take the approach of measuring the portion of the derived work that was based on copyrighted aspects of the original work, nor do we measure the simple distance between the works, nor the work required to generate those works.  Instead, we use a metric based on program description length which measures the advantage gained when building the allegedly-copied work both with and without access to the original work (see Section \ref{sec:definition}).  We argue that this method captures this ``importance,'' and we hope that this will enable a simpler ``comparison'' step of the test, freeing up the court's time to focus on the more challenging step of ``filtering.''
\section{Defining Derivation Similarity}
\label{sec:definition}

In this section, we describe our formal framework for modeling copyright questions, and we provide our mathematical test of \emph{derivation similarity} that can be used to help determine whether copyright infringement occurred.
To ensure that our eventual definition is built on solid computer science foundations, we begin in \S \ref{sub:formalism} with some formal definitions that measure the cost and complexity of clean room productions of the derived work.
Then, we provide our framework for reasoning about copyright infringement in \S \ref{sub:copyright-def}.
Finally, we discuss the strengths and limitations of this definition in \S \ref{sub:discussion}.

We emphasize that \S \ref{sub:formalism} is only needed to understand ``the math.'' Readers who are focused on the high-level conceptual ideas contained within our copyright infringement definition can safely skip directly to \S \ref{sub:copyright-def}.

\subsection{Some Computer Science Formalism}
\label{sub:formalism}

In this subsection, we include some formal definitions about the cost of a Turing machine and the complexity of a string.
This subsection is only needed to provide rigorous computer science foundations for the eventual copyright definition to follow in \S \ref{sub:copyright-def}.

We begin by formally defining the ``cost'' of a producer algorithm $\producer$ or reproducer algorithm $\reproducer$.
We assume throughout this work that all algorithms are deterministic.

\paragraph{Comparable}
We use $\original \simi \derived$ to denote that $\original$ and $\derived$ are \emph{comparable}.  In all cases we consider below, we define comparability using exact equality.
That said, we leave open the possibility that one might choose a different notion of comparable strings (e.g., being within a certain Hamming distance) depending on the situation in an individual case.
In the general case, we consider $\simi$ to be a binary relation on strings that need not be an equivalence relation; we only require that the relation is reflexive (i.e., every string is similar to itself).

\begin{definition}[Conditional Levin-Kolmogorov Cost]
\label{def:cost}
Let $\program$, $\inp$, and $\derived$ be strings in $\{0,1\}^\ast$.  Let $U(\program, \inp, t)$ be the output of a universal prefix Turing machine running program $\program$ for $t$ timesteps, where $\program$ has access to input tape $\inp$.  Then, the cost of running $\program$ on input \inp to get a string ``comparable'' to $\derived$ is
\begin{align*}
\dl{\program}{\simRep}{\inp} &= \min_{t \in \mathbb{N}} \{ \vert \program \vert + \lceil \log t \rceil : \simRep \gets U(\program, \inp, t) \} \\
\simdl{\program}{\derived}{\inp} &= \min_{\simRep \simi \derived} \{ \dl{\program}{\simRep}{\inp}\}
\end{align*}
\end{definition}

The first cost metric $\mathcal{C}$
is taken from the starting point of Levin's cost for universal search \cite{levin1973universal, DBLP:journals/iandc/Levin84} (which itself is a variant of Kolmogorov's complexity \cite{kolmogorov1963tables}) as presented in Def.\ 7.17 of Vity\'ani and Li \cite{DBLP:books/daglib/0071317}, and adding a conditional variant as used on pages 412-413 of the same.
Our new extended metric $\tilde{\mathcal{C}}$ provides additional flexibility by allowing for the possibility of (re)producing a similar string that is easier to generate.

In our analyses, we will also sometimes be interested in the \emph{minimum description length} of a program $\program$ that can output a string $\derived$, conditioned on some ``free inputs'' that $\program$ can freely use but is not charged any cost for receiving.
Importantly, this complexity metric is well-defined for a single string $\derived$, and it does not require the existence (much less an understanding of) a distribution from which $\derived$ is generated.

\begin{definition}[Conditional Levin-Kolmogorov Complexity]
\label{def:complexity}
Let $\inp$ and $\derived$ be strings in $\{0,1\}^\ast$.  Let $U(\program, \inp, t)$ be the output of a universal prefix Turing machine running program $\program$ for $t$ timesteps, where $\program$ has access to a tape $\inp$.  Then
\[
\levin{\derived}{\inp} = \min_{\program \in \{0,1\}^\ast} \simdl{\program}{\derived}{\inp}.
\]
\end{definition}

\subsection{Our Copyright Definition}
\label{sub:copyright-def}

In this subsection, we describe our abstract framework of the salient features in most copyright infringement cases, and we provide our main definition that separates the novel vs.\ copied creative expression in a derived work.
Our framework considers an original work $\original$ and derived work $\derived$, each of which contain 
copyrightable elements of creative expression.

Our objective is to compare the innate expression required to create the supposedly-infringing derived work $\derived$ both with and without access to the copyrightable elements of the original work $\original$.
In more detail, we want to understand the \emph{ex post} question about how ``simple'' it can be to create $\derived$ for a clean room actor working as effectively as possible, both with and without $\original$.
Our cost metric from \S \ref{sub:formalism} provides a way to do this.
To avoid reducing our definition to a mere measurement of the ``sweat of the brow,'' we also provide relevant facts and non-copyrighted works to the clean room actor for free.

We
formalize this definition in two ways:
as a standalone test that a court can apply on its own to determine the likelihood that copyright infringement occurred,
or (more interestingly) as an interactive game between the participants that is amenable to the adversarial system of justice used within courts in the United States.
In the latter viewpoint:
\begin{itemize}
\item The \plaintiff provides the smallest possible ``Producer'' algorithm \producer capable of independently producing something comparable to $\derived$, given full access to the original work $\original$ (and some other inputs we describe below).

\item The \defendant provides the smallest ``Reproducer'' algorithm \reproducer that produces a work comparable to $\derived$ \emph{without} using the copyrightable parts of $\original$, i.e., in the manner that the \defendant alleges that she initially created $\derived$.

\end{itemize}
By comparing the cost of $\producer$ and $\reproducer$, our definition will measure the extent to which the derived work $\derived$ intrinsically relies on the creative expression within $\original$,
as distinct from the independent expressive content that may also be present in $\derived$.

To focus on the copyrighted creative elements only, our framework also explicitly defines the following non-copyrighted materials such as basic facts or public domain works.
\begin{itemize}
\item The non-copyrightable aspects of the original work $\original$, which we denote $\noncopy(\original)$.

\item The non-copyrightable aspects of the derived work $\derived$.

\item A $\context$ comprising any other relevant works of interest to the $\plaintiff$ and $\defendant$.
\end{itemize}
We collectively refer to these non-copyrighted works as the background $\bg = (\noncopy(\original), \noncopy(\derived), \context)$.
Looking ahead, our definition will give $\producer$ and $\reproducer$ these background materials at zero cost when (re)producing the derived work $\derived$.

We require that the \plaintiff and \defendant agree upon the background material, before they attempt to use our definition.
This might sound counterintuitive: after all, how can the parties agree on the non-copyrighted aspects of $\original$ and $\derived$ when they are litigating whether $\derived$ itself infringes upon $\original$?

The idea here is that the $\plaintiff$ can scope her own claim: by conceding to the non-copyrighted aspects of $\original$ and $\derived$, the definition will hone in on the inherent expression contained within the remaining parts of $\original$ and $\derived$.
The $\plaintiff$'s goal is to identify a concrete portion of $\derived$ that bears a strong resemblance to (a portion of) $\original$, in the sense that a clean room implementation of $\derived$ would gain a substantial advantage by having access to 
the copyrightable parts of $\original$.
Conversely, the $\defendant$'s goal is to show that a clean room, armed with the background facts and works, could have created $\derived$ almost as easily with or without $\original$.
(See \S \ref{sub:discussion} for more details.)

Our formal definition simply takes the difference between these two costs.
We first present our empirical definition, which is an adversarial game between a \plaintiff who produces a single producer algorithm $\producer$ and a \defendant who produces a single reproducer algorithm $\reproducer$.
We call this notion \emph{derivation similarity} to indicate that it is a quantitative attempt at measuring substantial similarity based on the difference in $\producer$ and $\reproducer$'s ability to derive $\derived$ given their respective inputs.

\begin{definition}[Empirical Derivation Similarity]
\label{def:empirical}
Let $\original$ and $\derived$ denote the original and derived works, respectively. Given a producer algorithm $\producer$ and a reproducer algorithm $\reproducer$, the \emph{empirical derivation similarity} is defined as:
\begin{align*}
\DerSimEmp{\original}{\derived}{\producer}{\reproducer}{\bg}
= \simdl{\reproducer}{\derived}{\bg} - 
\simdl{\producer}{\derived}{(\bg, \original)},
\end{align*}
where $\bg = (\noncopy(\original), \noncopy(\derived), \context)$ contains the aspects of $\original$, $\derived$, and relevant other works that the parties agree are non-copyrighted.
\end{definition}

Our second definition is a standalone test that a court could apply on its own.
It assumes an ``optimal'' \plaintiff and \defendant who provide the lowest-cost $\producer$ and $\reproducer$ algorithms.

\begin{definition}[Theoretical Derivation Similarity]
\label{def:theoretical}
Let \original be the original work.
Let \derived be the derived work which the \plaintiff claims infringes on \original.
We compute the \emph{theoretical derivation similarity}
as:
\begin{align*}
\DerSim{\original}{\derived}
{\bg}
&=
\min_R \max_P \textsf{DerAdvEmp}(\original, \derived, \producer, \reproducer\,|\,
\bg) \\
&= \levin{\derived}{\bg)} - \levin{\derived}{(\bg, \original)}
\end{align*}
\end{definition}

\subsection{Discussion}
\label{sub:discussion}

In this subsection, we briefly describe several important concepts that influence the design of our mathematical definition as well as limitations of our framework.

\paragraph{Sliding scale of similarity}

Our definition purposely does not provide a binary `yes/no' test about whether copyright infringement occurred.
Instead, the output of $\Der$ is a number in between the following extremes:
\begin{itemize}
\item 0 if $\original$ and $\derived$ have no correlation.

\item $\levin{\derived}{\bg}$, the Levin-Kolmogorov complexity of the string $\derived$, if $\original = \derived$ (i.e., blatant copying occurred).
\end{itemize}
The goal here is to provide guidance to the courts to inform decisions about infringement.
We leave it to the courts to decide in any particular case what the cutoff points should be for determining that $\original$ and $\derived$ bear substantial similarity or striking similarity.
In our application of the definition to prior court cases in \S \ref{sec:precedent}, we describe why we think it is reasonable to conclude that $\Der$ is ``too large'' or ``small enough'' to conclude whether copyright infringement did or didn't occur.

\paragraph{Considerations when applying our definition}
Our framework has some limitations that influence whether and how it should be applied in court cases.

First, 
we assume that all relevant background information $\bg$ is available to both the \plaintiff and \defendant. As a result, our definition cannot be applied in cases where one party has trade secrets or other hidden information that would facilitate its clean room (re)production.

Second,
we remark that the $\context$ is vulnerable to ``gaming'' by a party to lower its conditional cost.
For instance, the \defendant might ask to include a 2-out-of-2 secret sharing of $\derived$ in the $\context$ (e.g., two strings $\derived_1$ and $\derived_2$ that individually look random and devoid of creative expression, but with $\derived = \derived_1 + \derived_2$) in order to reproduce $\derived$ at very low cost.
We rely on the adversarial system of justice to call out such abuses, and we stress that our definition only applies after consensus has been reached on the background $\bg$.
Generally speaking though, the intent is for $\context$ to contain the union of the \plaintiff and \defendant's desired prior works, as long as they are not ``gamed'' in this way.

Finally, we note that the theoretical derivation similarity $\Der$ is uncomputable in general, because the Levin-Kolmogorov complexity has this same defect.
For this reason, the main value of theoretical derivation similarity is in analyzing what we believe to be the likely outcome of court cases in our analysis in \S \ref{sec:precedent}-\ref{sec:synthetic}.
If this test is adopted by courts,
we recommend using empirical derivation similarity in practice.

\paragraph{Why minimum description length}

Our derivation similarity metric reasons about the ``amount of copying'' by comparing the (minimum) description length of the $\plaintiff$ and $\defendant$'s programs.
Here, we justify the use of a description-length metric as compared to other potential choices for the amount of copying, and then our specific choice of Levin-Kolmogorov complexity as a description-length metric.

Regarding alternative metrics: we chose to avoid measurements of cost based on running time because copyright doctrine is clear that the ``sweat of the brow'' necessary to generate the work is irrelevant to the question of copyright \cite[l. 359-360]{feist_v_rural}.
Another seemingly-natural choice for measuring the ``amount of copying'' in a work is mutual information, which measures how many bits of information the value of one random variable tells you about a second random variable.
The challenge is that mutual information applies to the probability distribution of all works that ``could'' have been created.
But, courts only have access to the actual strings $\original$ and $\derived$ rather than these counterfactual options, and even if convincing evidence could be provided about the distribution, there is not an established precedent for considering these counterfactual alternatives.
This also rules out other distribution-based tools, such as cryptographic notions of indistinguishability.

We also chose to avoid more naive notions of string comparison like edit distance, for two reasons.
First, edit distance is better suited to measuring the ``portion'' of the work that is copied, rather than its ``importance'' (see \S\ref{ssec:where-we-fit-in}).
Second, there is not a good way to incorporate the expression/idea dichotomy into these measures; that is, our framework must be able to identify that a large amount of expression is copied in a literally-copied novel, but a low amount of expression is copied in a literally-copied compilation of facts.  This property would also make it challenging to adjudicate correctly on merger doctrine cases  (see \S\ref{ssec:merger}).

Among description length metrics: an alternative natural choice is to consider Kolmogorov complexity, which removes the additive $\lceil \log t \rceil$ term from Def.~\ref{def:cost} and therefore only measures the size of the minimum-length Turing machine $M$.
We opt to use Levin complexity instead, for two reasons.
Philosophically, we want parties to provide an explicit description of their (re)production process;
we don't want a test that incentivizes the parties to perform 
a brute force search that simply tries all possible options until they get the right answer.
Practically, we don't want the existence of public-domain hash function digests to influence the substantial similarity test; see \S \ref{sub:hash} for details.

\paragraph{Numerical stability}

As with many computational, quantitative metrics, one challenge to consider when using our $\tilde{\mathcal{C}}$ cost metric is
numerical stability.
That is: if the strings $\original$ and $\derived$ both have very large Levin-Kolmogorov complexity, then it might be possible to ``sneak in'' extra effort into the programs $\producer$ or $\reproducer$ with very little net increase in their cost.
Put simply: our derivation similarity metric doesn't work well when the costs on both sides are large.

Numerical instability harms the $\plaintiff$, since by construction the $\defendant$'s task in Def.~\ref{def:empirical} is necessarily harder and it is the $\plaintiff$'s goal to ensure that this difference is noticeable.
To resolve the numerical stability issue, we rely on the $\plaintiff$'s ability to declare aspects of $\original$ as non-copyrighted and thus available ``for free'' to $\producer$ and $\reproducer$, even materials that might actually be copyrighted in practice, to scope the copyright claim down to the (hopefully small amount of) infringing material and mitigate this numerical stability concern.

\section{Adherence to Precedent}
\label{sec:precedent}

In this section, we apply our framework to several prominent historical copyright cases, both to demonstrate that our framework reaches the same result as the courts in all cases, and also to illustrate how reasoning with the framework works.  We begin with cases that illustrate the \emph{idea vs.\ expression} dichotomy in \S\ref{ssec:idea-expression}, which in our framework concerns what material belongs in $\noncopy(\original)$.  We then proceed in \S\ref{ssec:merger} to analyze cases which illustrate our framework's approach to the \emph{merger doctrine} --- our framework does not require applying any special reasoning in this case; the best $\producer$ and $\reproducer$ in these cases will simply operate via reasoning that resembles the merger doctrine.  We end by considering cases that use computer code and describing our framework's relationship to the \emph{Abstraction-Filtration-Comparison} framework, in \S\ref{ssec:computer-code}.

\subsection{Idea vs.\ Expression}
\label{ssec:idea-expression}

\paragraph{Feist Publications, Inc.~v.~Rural Telephone Service Co., 499 U.S. 340 (1991)}

\emph{Feist} \cite{feist_v_rural} famously resolved the seeming tension arising from the notion that facts are not copyrightable, but compilations of facts can be.  It also corrected a misconception in lower courts by stating that copyright is \emph{not} a reward for the ``sweat of the brow'' that went into creating the work.  Rather, copyright only extends to \emph{original} works, meaning those that are ``independently created by the author, and possess[] some minimal degree of creativity'' (\cite[l. 345]{feist_v_rural} citing \cite[Vol. 1 \S2.01]{nimmer}) and ``[n]o one may claim originality as to facts'' (\cite[l. 347]{feist_v_rural} citing \cite[Vol 1 \S2.11]{nimmer}).  Thus a compilation of facts may be copyrightable, but only if they are arranged or selected in an original manner, making this protection ``thin'' \cite[l. 349]{feist_v_rural}.

In this famous case, Feist Publications published a telephone directory that used several thousand entries of an existing copyrighted phone book, Rural Telephone Service Co., including 1,309 identical entries out of 46,878.  The court found no ``copying of constituent elements of the work that are original'' \cite[l. 361]{feist_v_rural}, since neither the raw facts themselves, nor Rural's arrangement and selection of those facts (name, town, and telephone number of each person in Rural's telephone service, listed alphabetically by last name), were original.

Recall that our framework compares the size of the smallest possible $\producer$ and $\reproducer$ that generate works similar to $\derived$ --- in this case, we take our similarity metric $\simi$ to be exact equality.  Recall that $\producer$ has access to all of $\original$, and $\reproducer$ has access only to the noncopyrightable aspects of $\original$.
In modeling the facts of \emph{Feist}, because the facts and their selection and arrangement in Rural's phone book were not original, nearly the entire text of Rural's phone book is a part of $\noncopy(\original)$.  
Feist also copied four fictitious listings that Rural had included expressly to detect copying;  although the court in \emph{Feist} did not address this matter directly, most courts treat fictitious facts as facts and do not afford them copyright protections (see e.g. \cite{davies_v_bowes}, though this approach has not been applied perfectly consistently \cite{smith2019truth}).
In our framework we treat these as elements of $\noncopy(\original)$, but not $\context$.

The court did not rule on whether or not Feist's phone book is copyrightable --- again, neither the facts themselves nor the arrangement of those facts were original, but since Feist had more freedom in how it chose which entries to include in its book in the first place.  (Rural was a monopoly telephone provider in a local area and its phone book included only the phone numbers it provided plus the small number of fictitious entries --- about 7,700 total.  Feist covered 11 telephone service areas across 15 counties, for about 47,000 total.)
To model this, as shown in Table~\ref{tab:feist_v_rural}, we could consider $\context$ to contain a much larger collection of facts arranged and chosen in an un-original way --- perhaps every phone number, name, and address in the state (the sorting mechanism does not matter, it will change \producer and \reproducer's size by the same amount, but it is likely that sorting by county and then alphabetically will yield the smallest results).  $\noncopy(\derived)$ contains a list of the counties from which the   In order to select the facts to go into the Feist phone book, the $\producer$ and $\reproducer$ must generate the set of indices within $\context$ of the entries in Feist's phone book.  Crucially, both $\producer$ and $\reproducer$ must do this --- nothing in $\copyr(\original)$ aids this process.

Thus, the difference in description length between $\reproducer$ and $\producer$ is  low and it is reasonable to conclude that no copyright infringement occurred.

\begin{table*}
\small
\begin{tabular}{|p{0.49\textwidth}|p{0.49\textwidth}|} \hline
\multicolumn{2}{|p{0.98\textwidth}|}{\tabtitle{Encoding} Assuming exact equality required.  All entries converted to raw text, since aspects of the phone book such as the font or physical properties were not in dispute.} \\ \hline
\tabtitle{$\original$} Rural's phone book & 
\tabtitle{$\noncopy(\original)$} All entries in Rural's phone book, in alphabetical order \\ \hline
\tabtitle{$\derived$} Feist's phone book & 
\tabtitle{$\noncopy(\derived)$} The list of counties covered in Feist's phone book \\ \hline
\multicolumn{2}{|p{0.98\textwidth}|}{\tabtitle{\context} All phone numbers, addresses, and names in the state, sorted by county then by last name} \\ \hline
\multicolumn{2}{|p{0.98\textwidth}|}{\tabtitle{Notes} Let $L_\context$ be a compressed list of indices of phone numbers in the context in $\derived$, with null characters separating the entries in uncompressed form.  
Let $L_\original$ be a similar alphabetized list of entries of entries within $\original$ (both real and fictitious), which as discussed are part of $\noncopy(\original)$.
} \\
\hline \hline
\tabtitle{Best $\producer$} $\producer$ must include the smallest possible compression $L_\context$ and $L_\original$.  Also includes code of constant length $c$ to fuse the entries from $\context$ and $\noncopy(\original)$ denoted in those lists, and write them in the output $\simProd$ in alphabetical order.  This should take $|L_\context| + |L_\original|$ time.& 
\tabtitle{Best $\reproducer$} $\reproducer$ must include $L_\context$ and $L_\original$ similar to $\producer$.  Also includes code $c'$ to copy the information from $L_\context$ and $L_\original$ into $\simRep$ in alphabetical order.  Note that $c \approx c'$.  The copying should take approximately $|L_\context| + |L_\original|$ time. \\ 
\tabtitle{$\levin{\derived}{(\bg, \original)}$} $|L_\context| \log|L_\context| + |L_\original|\log|L_\original| + c + \log(|L_\context| + |L_\original|)$ & 
\tabtitle{$\levin{\derived}{\bg}$} $|L_\context| \log |L_\context| + |L_\original| \log |L_\original| + c' + \log(|L_\context| + |L_\original|)$ \\ \hline
\multicolumn{2}{|p{0.98\textwidth}|}{\tabtitle{Outcome} $\copyclose{\original}{\derived}{\bg} = (c'-c) \approx 0$.}  \\
\multicolumn{2}{|p{0.98\textwidth}|}{\tabtitle{Court outcome} No infringement occurred.} \\ \hline
\end{tabular}
\caption{Our framework's treatment of \emph{Feist v. Rural}}
\label{tab:feist_v_rural}
\end{table*}

\paragraph{Baker v.~Selden, 101 US 99 (1880)}

In this case \cite{baker_v_selden}, Selden had published a book entitled \emph{Selden's Condensed Ledger, or Book-keeping simplified} and several other books on the topic of his book-keeping method.  He alleged that Baker infringed on those books' copyright by publishing his own book describing a similar book-keeping method, and especially using similar illustrations to describe the technique.

This case is key in describing the difference between copyrighting an expression of the system versus the system itself.  They state ``By publishing the book, without getting a patent for the art [the system], the latter is given to the public.  The fact that the art described in the book by illustrations of lines and figures which are reproduced in practice in the application of the art, makes no difference.  Those illustrations are the mere language employed by the author to convey his ideas more clearly.  Had he used words of description instead of diagrams (which merely stand in place of words), there could not by the slightest doubt that others, applying the art to practical use, might lawfully draw the lines and diagrams which were in the author's mind, and which he thus described by words in his book" \cite[l. 103]{baker_v_selden}.
Our analysis of this case, shown in Table~\ref{tab:baker_v_selden}, describes how since although the book-keeping method described in $\derived$ is similar to the one described in $\original$, once those book keeping methods are provided separately in $\noncopy(\derived)$ and $\noncopy(\original)$, the remaining expression in $\original$ does not aid the creation of $\derived$.

\begin{table*}
\small
\begin{tabular}{|p{0.49\textwidth}|p{0.49\textwidth}|} \hline
\multicolumn{2}{|p{0.98\textwidth}|}{\tabtitle{Encoding} Text and pictoral snapshots of the descriptions of each book and essay.  Similarity taken to be exact equality.} \\ \hline
\tabtitle{$\original$} The exact text of ``an introductory essay explaining the book-keeping system [in Selden's book], to which are annexed certain forms or blanks, consisting of ruled lines, and headings'' \cite[l. 100]{baker_v_selden} & 
\tabtitle{$\noncopy(\original)$} The book-keeping methods described in $\original$. \\ \hline
\tabtitle{$\derived$} A series of books that describes a book-keeping method ``similar \ldots so far as results are concerned, but makes a different arrangement of the columns, and uses different headings'' \cite[l. 100]{baker_v_selden} & 
\tabtitle{$\noncopy(\derived)$} The book-keeping methods described in $\derived$ \\ \hline
\multicolumn{2}{|p{0.98\textwidth}|}{\tabtitle{\context} An unrelated public domain book describing a book-keeping system.} \\ \hline
\multicolumn{2}{|p{0.98\textwidth}|}{\tabtitle{Notes} Let $s(\bg)$ be the smallest possible algorithm which reads in the book-keeping methods in $\noncopy(\original)$ and $\noncopy(\derived)$, and reads in the related book from $\context$, and outputs $\derived$.  Let $s'(\bg, \original)$ be the smallest possible algorithm which reads $\original$ in addition to the other inputs above. By assumption, conditioned on $\bg$, $\derived$ and $\original$ should not be much more similar than $\derived$ and the unrelated book in $\context$ aside from the actual book-keeping method described, which is fully contained in $\noncopy(\original)$ and therefore accessible to both $\producer$ and $\reproducer$.  Thus, we expect $|s| \approx |s'|$ and we expect their runtimes $t$ and $t'$ to be similar as well.} \\
\hline \hline
\tabtitle{Best $\producer$} Runs $s(\bg)$ & 
\tabtitle{Best $\reproducer$} Runs $s'(\bg, \original)$ \\ 
\tabtitle{$\levin{\derived}{(\bg, \original)}$} $|s| + \log t$ & 
\tabtitle{$\levin{\derived}{\bg}$} $|s'| + \log t'$ \\ \hline
\multicolumn{2}{|p{0.98\textwidth}|}{\tabtitle{Outcome} $\DerAdv{\original}{\derived}{\bg} = |s| + \log t - (|s'| + \log t') \approx 0$} \\
\multicolumn{2}{|p{0.98\textwidth}|}{\tabtitle{Court outcome} No infringement occurred.} \\ \hline
\end{tabular}
\caption{Our framework's treatment of \emph{Baker v. Selden}}
\label{tab:baker_v_selden}
\end{table*}

\paragraph{Southco, Inc. v. Kanebridge Corp., 390 F.3d 276 (2004)}

In this case, Southco, Inc. had created a part numbering system to uniquely identify and describe characteristics of parts like screws.  Kanebridge Corp. had labeled its screws both with its own part numbering system, and with Southco's numbering system, to describe that the parts were interchanbeable.  
The court ultimately found that the part numbers are not copyrightable, for two reasons.  First, Southco had not claimed any copyright on the numbering \emph{system} --- in our framework, the system is in the context.  And once the system is fixed, no creative options remain in picking the numbers for a particular screw --- in our framework, $\noncopy(\original) = \noncopy(\derived)$ contains a physical description of the screw, and $\original = \derived$ is the part number for the screw itself, which could have been found by ``looking up'' the physical description in the parts numbering system.  The second reason the numbers were not copyrightable was that they are akin to short phrases or titles, which were determined not to be copyrightable.  We believe this to essentially set a bar for how much bigger \reproducer is allowed to be than \producer --- if the difference involves a small amount of code to generate a ``phrase,'' no copyright infringement has occurred.

\subsection{The Merger Doctrine}
\label{ssec:merger}

\paragraph{Herbert Rosenthal Jewelry Corp. v. Kalpakian, 446 F.2d 738 (9th Cir. 1971)}

Herbert Rosenthal Jewelry Corp. designed and copyrighted a jeweled bee-shaped pin, and Edward and Lucy Kalpakian had ``manufactured a jeweled bee alleged to be similar'' \cite{rosenthal_v_kalpakian}.  The plaintiff especially focused on the arrangement of jewels on the original bee pin.  However, the plaintiff was unable to describe exactly which aspects of the arrangement were original, nor what alternative arrangements would not infringe the copyright.  The defendant's arrangement was ``simply a function of the size and form of the bee pin and the size of the jewels used.''

In our framework, this ``function'' can be done explicitly by the Reproducer --- this can be thought of as a bin-packing problem, fitting jewels of different sizes into rows across the bee's surface.   Note that even if the Kalpakians'  arrangement of jewels had been identical to Rosenthal's, the maximum difference in the size difference between \reproducer and \producer is the size of the algorithm necessary to produce that arrangement. 

\begin{table*}
\small
\begin{tabular}{|p{0.49\textwidth}|p{0.49\textwidth}|} \hline
\multicolumn{2}{|p{0.98\textwidth}|}{\tabtitle{Encoding} Structured lists of jewel arrangements by type, size, arrangement, and color.  Exact equality required.} \\ \hline
\tabtitle{$\original$} The arrangement of 19 white jewels on Herbert Rosenthal Jewelry Corp's jeweled bee & 
\tabtitle{$\noncopy(\original)$} The size and shape of $\original$ \\ \hline
\tabtitle{$\derived$} A set of additional jeweled bees ``in three sizes decorated with from nine to thirty jewels of various sizes, kinds, and colors'' \cite[l. 740]{rosenthal_v_kalpakian} & 
\tabtitle{$\noncopy(\derived)$} The size and shape of $\derived$ \\ \hline
\multicolumn{2}{|p{0.98\textwidth}|}{\tabtitle{\context} A list of potential jewel colors and sizes} \\ \hline
\multicolumn{2}{|p{0.98\textwidth}|}{\tabtitle{Notes} Let $A$ be a compressed pattern of jewels on $\derived$. Let $O$ be a compressed list of the indices of jewels that overlap between $\original$ and $\derived$.  Conditioned on the idea of a jeweled bee, $O$ is small as described by the court \cite[l. 742]{rosenthal_v_kalpakian}. We assume that the rumtime of algorithms placing the pattern $A$ is approximately $|A|$ regardless of whether $O$ is used or not.}  \\
\hline \hline
\tabtitle{Best $\producer$} The pattern of jewels on $\derived$ must be hardcoded; in areas where there are overlaps with $\original$, a strcopy can be used. & 
\tabtitle{Best $\reproducer$} The same pattern must be hardcoded, the only difference is that any overlaps with $\original$ cannot use copy. \\ 
\tabtitle{$\levin{\derived}{(\bg, \original)}$} $|A| - |O| + \log |A|$ & 
\tabtitle{$\levin{\derived}{\bg}$} $|A| + \log |A|$ \\ \hline
\multicolumn{2}{|p{0.98\textwidth}|}{\tabtitle{Outcome} $\DerAdv{\original}{\derived}{\bg} = |O| \approx 0$ by assumption}  \\
\multicolumn{2}{|p{0.98\textwidth}|}{\tabtitle{Court outcome} No infringement occurred.} \\ \hline
\end{tabular}
\caption{Our framework's treatment of \emph{Rosenthal v. Kalpakian}}
\label{tab:rosenthal_v_kalpakian}
\end{table*}

\paragraph{Morrissey v. Procter \& Gamble Company, 379 F.2d 675 (1st Cir. 1967)}

In \emph{Morrissey v. Proctor \& Gamble Co.} \cite{morrissey_v_pg}, Morrissey had written a promotion for a sweepstakes, and Proctor \& Gamble Co. had later released a similar sweepstakes with some very similar text.  

The court ruled that although Morrissey's expression did contain ``original creative authorship,'' and ``there was more than one way of expressing even this simple substance'' \cite[l. 678]{morrissey_v_pg}, the ``uncopyrightable subject matter is very narrow, so that the topic necessarily requires ... if not only one form of expression, at best only a limited number'' then copyright does not apply.

This would come to be a widely cited case establishing that the merger doctrine could apply to a case where there is more than one way to express some idea, but still only a limited number of ways to express the idea that are at least as reasonable \cite{samuelson2016reconceptualizing}.

Our framework is well-suited to handle the merger doctrine in this form.  If there are only a limited number of ways to represent an idea, then those ideas may be enumerated by both \producer and \reproducer using only a very short code description, or possibly even enumeration in the \context, as appropriate.  Moreover, our framework points to drawing a potential line of \emph{how many} ways to represent an idea should be acceptable, since merger doctrine does not have to be treated any differently under our framework than any other copyright case.

\subsection{Computer Code}
\label{ssec:computer-code}

There have been a number of rulings on what aspects of computer code are and are not copyrightable.  \emph{Apple computer, Inc. v. Franklin Computer Corp.)} \cite{apple_v_franklin} stated that object code (as opposed to source code) is copyrightable, though their rationale that copyright extends to works from which the expression ``can be perceived, reproduced, or otherwise communicated, either directly or with the aid of a machine or device'' may fall flat for obfuscated code.
\emph{Lotus Dev Corp. v. Borland Int'l} \cite{lotus_v_borland} additionally stated that menu hierarchies are not copyrightable,
and \emph{Sega Enterprises Ltd. v. Accolade, Inc.} \cite{sega_v_accolade} decided that dissasembly can sometimes fall under fair use, if the dissasembly only ``reveals'' un-copyrighted parts of the code and there is a ``legitimate reason'' for accessing this.

We examine some other cases in more detail.

\paragraph{Computer Associates International, Inc. v. Altai, Inc., 982 F.2d 693 (1992)}

\emph{Altai} \cite{ca_v_altai} introduced the \emph{Abstraction-Filtration-Comparison} framework for determining whether computer programs are substantially similar.  The test applies to the \emph{original} work, not to the derived work.  The test first describes the original program at different levels of ``abstraction'' (``At low levels of abstraction, a program's structure may be quite complex; at the highest level it is trivial'' \cite[l. 707]{ca_v_altai}).  Then, a ``successive filtering method'' \cite[l. 707]{ca_v_altai} separates copyrightable expression from non-copyrightable elements.  For instance, elements dictated by efficiency or external factors (such as hardware specifications or compatibility requirements), or elements in the public domain, are removed.  After this, whatever potentially-copyrightable elements remain of the original program are compared to the derived program.

Computer Associates (CA) sold a program entitled CA-SCHEDULER which was a scheduler for IBM mainframe computers.  One component of CA-SCHEDULER was an interface called ADAPTER; instead of making system calls to the operating system directly, CA-SCHEDULER would call ADAPTER, which would correctly call whichever of the three possible operating systems used by IBM mainframe computers the main program was being run on.
Altai, Inc. had a scheduler called ZEKE, which Altai wished to rewrite for use in a particular operating system.  An employee at Altai offered a job to an employee of CA, Arney, to help write the new ZEKE version.  When Arney joined Altai in 1984, he brought source code for ADAPTER, ``in knowing violation of the CA employee agreements he had signed'' \cite[l. 700]{ca_v_altai}.  He also suggested that the best way to write multiple versions of ZEKE for different operating systems would be to have them call a common interface, without revealing that this idea stemmed from ADAPTER.  Altai's version of the interface was called OSCAR, and ultimately about 30\% of its code was copied from ADAPTER.
A few years later, CA learned that much of the OSCAR code was copied, from ADAPTER, and contacted Altai.  The company excised the ADAPTER-copied code and set several programmers who had not been on the OSCAR program before to rewrite it, excluding Arney from the process.
The initial copying of ADAPTER into OSCAR was not in dispute.  But CA continued to accuse the rewritten OSCAR of infringing the CA-SCHEDULER code for copying its ``non-literal elements'' \cite[l. 701]{ca_v_altai}.
The court ultimately rejected CA's argument.  After abstracting ADAPTER and filtering out the non-copyrightable elements, the code ``presented no similarity at all'' \cite[l. 714]{ca_v_altai}.  The main elements at issue ended up being external constraints to ensure compatibility with the operating system.

Our framework essentially formalizes the latter two steps of the Abstraction-Filtration-Comparison framework.  We also rely on manual abstraction or encoding of specific elements, and then filter the non-copyrightable elements into $\noncopy(\original)$.  However, rather than ``comparing'' the remaining copyrightable info from $\original$ and $\derived$, we use a metric of description length to measure this instead.
In our framework, $\original$ is the ADAPTER program and $\noncopy(\original)$ is the portions of ADAPTER that were made necessary by compatibility with the operating system.  $\derived$ and $\noncopy(\derived)$ are similarly defined for the rewritten OSCAR program.  The \context contains a description of the operating system calls that both programs must adhere to.  

\begin{table*}
\small
\begin{tabular}{|p{0.49\textwidth}|p{0.49\textwidth}|} \hline
\multicolumn{2}{|p{0.98\textwidth}|}{\tabtitle{Encoding} Computer code} \\ \hline
\tabtitle{$\original$} The ADAPTER program &
\tabtitle{$\noncopy(\original)$} Those parts of ADAPTER made necessary by compatibility with the operating system \\ \hline
\tabtitle{$\derived$} The OSCAR program & 
\tabtitle{$\noncopy(\derived)$} Those parts of OSCAR made necessary by compatibility with the operating system \\ \hline
\multicolumn{2}{|p{0.98\textwidth}|}{\tabtitle{\context} A description of the operating system calls that both programs must adhere to} \\ \hline
\multicolumn{2}{|p{0.98\textwidth}|}{\tabtitle{Notes} By the court's determining, all parts of $\derived$ were either within $\noncopy(\original)$ or were dissimilar to $\original$ \cite[l. 714]{ca_v_altai}.  Let $C$ be the smallest possible program that takes $\context$ and $\noncopy(\original)$ as input, and outputs $\derived$.  By assumption, $C$ is very similar to the smallest possible $C'$ that creates $\derived$ taking $\context$ and $\original$ as input.  We assume they have similar sizes ($s$ and $s'$) and runtimes ($t$ and $t'$).} \\
\hline \hline
\tabtitle{Best $\producer$} Runs $C'$ &
\tabtitle{Best $\reproducer$} Runs $C$ \\ 
\tabtitle{$\levin{\derived}{(\bg, \original)}$} $s'$+$\log t'$ & 
\tabtitle{$\levin{\derived}{\bg}$} $s+\log t$ \\ \hline
\multicolumn{2}{|p{0.98\textwidth}|}{\tabtitle{Outcome} $\DerAdv{\original}{\derived}{\bg} = s + \log t - (s' + \log t') \approx 0$}  \\
\multicolumn{2}{|p{0.98\textwidth}|}{\tabtitle{Court outcome} No infringement occurred.} \\ \hline
\end{tabular}
\caption{Our framework's treatment of \emph{Computer Associates v.~Altai}}
\label{tab:ca_v_altai}
\end{table*}

\section{Testing the limits with hypothetical examples}
\label{sec:synthetic}
\label{ssec:crypto}

In this section, we describe a few synthetic scenarios with new wrinkles that go beyond the precedent-setting cases from \S \ref{sec:precedent}.
Our objective in this section is to elucidate and stress-test elements of our definition, in order to further justify some of our modeling decisions.

\subsection{Arrangements of non-original fragments can be original}

In \emph{Knitwaves v. Lollytogs}, the Second circuit stated that if they had to compare each individual element of the work instead of examining the work's ``total concept and feel'' \cite[l. 1003]{knitwaves_v_lollytogs}, then taken ``to its logical conclusion, [the court] might have to decide that `there can be no originality in a painting because all colors of paint have been used somewhere in the past.''' \cite[l. 1003]{knitwaves_v_lollytogs}.

As the court goes on to point out \cite[l. 1004]{knitwaves_v_lollytogs}, the fact that a work is composed of uncopyrightable elements does not preclude copyright over the \emph{compilation} of those elements, as long as that compilation contains originality \cite[l. 344]{feist_v_rural}.  
In some sense any work can be broken down in this way --- novels are ``just'' compilations of words or letters, digital images are ``just'' collections of pixels, and music is ``just'' arrangements of notes.

Thus, it is important that our framework ensures that, for example, representing a novel as ``a collection of words'' does not yield a smaller program $\producer$ than hard-coding the novel itself does --- we wish to ensure there is no mysterious advantage gained in representing a novel just because one has access to an unrelated reference text like a dictionary --- or any other book, for that matter.

Suppose $\producer$ naively read an index into the dictionary each time it needed to use a word.  Depending on which words one chose to include, there are estimated to be between 400,000 and 1,000,000 words in the English dictionary, i.e. it requires approximately 19-20 bits to represent a word by its index into a dictionary in the $\context$. Shannon used Zipf's law to estimate the entropy of an English word at approximately 11.81 bits \cite{shannon1951prediction}, and this can be reduced with an improved estimate of the distribution of words \cite{GRIGNETTI1964304}.

In some sense this comparison is unfair to the dictionary approach --- compression schemes use fewer bits to represent more frequently occurring words --- in fact many compression schemes store their own dictionaries in this way \cite[ch. 5]{sayood2017introduction}; in contrast our naive dictionary approach insists upon using the log of the number of entries in the dictionary to index within it.
But the point is that the compression will not be improved by a dictionary that is not specific to the work in question.  Dictionaries arranged in more advantageous ways compared to $\derived$ can be considered part of the compression of $\derived$ itself, and so no longer belong in the context.

\subsection{Hash functions cannot be used as a shortcut}
\label{sub:hash}

We imagine a setting where a plaintiff has copyrighted some work $\original$, but places the hash $h(\original)$ of the work in the public domain.
This scenario is not theoretical; hashes of potentially-copyrighted files are often used in downloads to ensure that the file being downloaded is the correct one --- the hash of the downloaded file is compared to a published hash to ensure that no corruption or tampering occurred during the download.

We imagine that $h$ is a truly random function $h : \{0,1\}^\ast \rightarrow \{0,1\}^n$.  (This assumption makes the analysis easier, and we do not expect our results to change meaningfully with a different choice of hash function model.)  With overwhelming probability, each image will have infinitely many preimages.  A machine that brute-forces messages in lexicographic order until it finds \emph{a} preimage of $h(\original)$ is quite small.  However, being able to specify \emph{the right} preimage (that is, $\original$), would still require a significant description length, as we proceed to show.

Suppose we consider the binary length of the work $\ell$ to be part of $\noncopy(\original)$, and that $h$ hashes strings of that length (i.e., $h : \{0,1\}^\ell \rightarrow \{0,1\}^n$).  We expect approximately $2^{\ell-n}$ possible preimages to hash to $h(\original)$, assuming $\ell \gg n$.  Indexing within this set does not save much space compared to needing to write out the entirety of $\original$ anyway. The set (and thus indices into it) remain large even when restricted to, e.g., strings with proper English grammar.

This shows that a pure description-length framework does not lose ``much'' derivation similarity by placing a hash of the copyrighted work in the public domain.
Nonetheless, this example motivates us to adopt the Levin-Kolmogorov description length framework we defined formally in \S\ref{sec:definition} rather than a pure Kolmogorov definition, for two reasons.

The first reason is non-technical: it is our intent for the Producer and Reproducer to bear some resemblance to the actual method an author would use when creating the work $\derived$ from the various inputs.  Running the program of ``brute force search through the options until locating one that looks like $\derived$'' does not meet this goal.  This motivates the use of Levin-style complexity measures that were originally created in the exact context of brute-force search \cite{levin1973universal, DBLP:journals/iandc/Levin84}.

The second reason is technical: we do not wish the brute-forcing method to aid $\producer$ \emph{even a little bit}.  Thus, small reduction in size achieved by doing a brute-force search for a certain element is offset by charging for the brute-force search (the log of the machine's runtime).
This has virtually no impact on efficient machines, and only becomes important in the setting of a machine as inefficient as brute-force search.

\section{Zero-knowledge Proofs of Derivation Similarity}
\label{sec:zk}

In this section we consider the following question: what if the \plaintiff and \defendant want to submit to the court the result of an empirical derivation similarity test, but do not wish to reveal the producer program $\producer$ and/or reproducer program $\reproducer$?
This question is inspired by Bamberger et al.~\cite{verification_dilemmas}, who discuss the dilemma that occurs when
``a certain computer program $\producer$\ldots is the plaintiff's trade secret''
and still the goal is that
``the parties can determine similarity of the programs while keeping them secret.''
Their work pointed out the opportunity to use zero knowledge proofs in cases involving trade secrets.
Our goal here is to highlight the value of zero knowledge in copyright cases as well, where some of the same dynamics may apply.

\paragraph{Value to the legal system}
Before delving into technical details, we make three remarks about why it might be valuable from a legal perspective to reveal only the result of a derivation similarity test along with a proof of correctness, and why all parties might have incentives to participate.

Our primary notivation is the same as Bamberger et al.~\cite{verification_dilemmas}: we are concerned about the possibility that the \plaintiff and \defendant might be less likely to litigate a copyright case if 
the (re)producer algorithms $\producer$ and $\reproducer$ reveal some ``secret sauce'' about the creativity in their methods.
Our framework does not require $\producer$ and $\reproducer$ to be publicly visible (only that the background information $\bg$ is available to both parties),
so pursuing a privacy-preserving test is compatible with the concepts discussed in \S \ref{sub:discussion}.

Second, even though we have focused in this work on litigation in the court system, we envision a private derivation similarity test to be a useful first step in private mediation proceedings.
If these proceedings break down and parties move their case to the courtroom, then only a zero knowledge proof needs to be placed in the public record.

Third, even the ``losing'' party of our derivation similarity test may participate in its computation, both because it may not yet know the final result, and because the party has ample recourse to contest a copyright decision on other grounds even if the similarity metric is not in their favor.
Recall that the goal of derivation similarity is to provide a predictable test of where the parties stand with respect to substantial similarity, precisely to allow the parties and courts to focus their litigation on other aspects of copyright law like fair use, DMCA anti-circumvention rules, whether the original work $\original$ is copyrightable in the first instance, and so on.

\paragraph{Privacy-preserving derivation similarity}
From a computer science standpoint, the ideal tool to use here is \emph{secure multiparty computation} \cite{mpc-book}, which allows two parties to perform a joint calculation based on hidden data while learning no more than (what can be inferred from) the result.
In this case, the \plaintiff has a secret program $\producer$, the \defendant has a secret program $\reproducer$, both parties know $(\original, \derived, \bg)$,
and they want to calculate the empirical derivation similarity $\DerSimEmp{\original}{\derived}{\producer}{\reproducer}{\bg}$.

Observe that the empirical derivation similarity is the difference of two numbers that $\plaintiff$ and $\defendant$ individually know.
As a consequence, given only the result of the $\DerEmp$ calculation, each party will learn the Levin-Kolmogorov 
cost 
of the other party's program.
Put another way: there is no point in trying to hide the 
$\tilde{\mathcal{C}}$
metrics of each party from each other, even though there may be value in hiding this information from the general public.

Consequentially, there is a simple design for a secure computation algorithm for $\DerEmp$: each party constructs a 
\emph{zero knowledge proof} of the complexity of their own program.
Concretely, the \plaintiff publishes the claimed cost 
$c_\producer = \simdl{\producer}{\derived}{(\bg, \original)}$,
a (possibly very loose) upper bound $T$ on the runtime of $\producer$, and
a zero knowledge proof of knowledge of the statement ``I know a producer program $\producer$, a string $\simRep$, and a time bound $t$ such that (i) $\simRep \simi \derived$, (ii) $U(\program, (\bg, \original), t) = \simRep$ where $U_T$ is a universal Turing machine running in $T$ steps, and (iii) $c_\producer = \vert \producer \vert + \lceil \log t \rceil$.''
Similarly, the \defendant proves in zero knowledge that they know $\reproducer$ that can create something similar to $\derived$ with $c_\reproducer$ Levin-Kolmogorov cost, but without free access to $\original$.
Then, the parties work together to produce a recursive zero knowledge proof about the value of $c_\reproducer - c_\producer$ while hiding the two values individually.

Using modern zero knowledge arguments, the resulting proofs, to be placed in the public record can be a few hundred kilobytes in size (e.g., using \cite{EC:Groth16}). Furthermore, using  incrementally-verifiable computation  \cite{Valiant08inc,STOC:BCCT13,BCTV14scalable}, the time to produce the proofs is linear in running time of $\producer$ and $\reproducer$.
This proof will hide the parties' programs while also allowing for public verifiability of both the result and the process of executing the derivation similarity test.

\section*{Acknowledgments}

We thank Ed Felten and Stacey Dogan for their helpful discussions and valuable insights about computer science and intellectual property law.
Sarah Scheffler was supported by a Google Ph.D.\ Fellowship and the Center for Information Technology Policy at Princeton University.
Eran Tromer was supported by the DARPA SIEVE program under Contract No.\ HR001120C00.
Mayank Varia was supported by the National Science Foundation under Grants No.\ 1718135, 1801564, 1915763, and 1931714,
by the DARPA SIEVE program under Agreement No.\ HR00112020021,
and by DARPA and the Naval Information Warfare Center (NIWC) under Contract No.\ N66001-15-C-4071.
Any opinions, findings and conclusions or recommendations expressed in this material are those of the author(s) and do not necessarily reflect the views of
Google, CITP, DARPA, NSF, NIWC, or any other organization.

\bibliography{bib-extra,bib-abbrev3,bib-crypto}

\begin{thebibliography}{10}

\bibitem{17usc102}
{17 U.S. Code \S 102 - Subject matter of copyright: In general}.

\bibitem{17usc106}
{17 U.S. Code \S 106 - Exclusive rights in copyrighted works}.

\bibitem{copyright76}
{94th United States Congress}.
\newblock {Public Law 94-553 -- Copyright Act of 1976}.
\newblock \url{http://uscode.house.gov/statutes/pl/94/553.pdf}, October 1976.

\bibitem{AbelsonABBBDGG015}
Harold Abelson, Ross~J. Anderson, Steven~M. Bellovin, Josh Benaloh, Matt Blaze,
  Whitfield Diffie, John Gilmore, Matthew Green, Susan Landau, Peter~G.
  Neumann, Ronald~L. Rivest, Jeffrey~I. Schiller, Bruce Schneier, Michael~A.
  Specter, and Daniel~J. Weitzner.
\newblock Keys under doormats: mandating insecurity by requiring government
  access to all data and communications.
\newblock {\em J. Cybersecur.}, 1(1):69--79, 2015.

\bibitem{Altman2021hybrid}
Micah Altman, Aloni Cohen, Kobbi Nissim, and Alexandra Wood.
\newblock What a hybrid legal-technical analysis teaches us about privacy
  regulation: The case of singling out.
\newblock {\em BU J Sci. \& Tech. L.}, 27:1, 2021.

\bibitem{apple_v_franklin}
{Apple Computer, Inc. v. Franklin Computer Corp., 714 F.2d 1240 (3d Cir.
  1983)}.

\bibitem{arnstein_v_porter}
{Arnstein v. Porter, 154 F.2d 464 (2d Cir. 1946)}.

\bibitem{baker_v_selden}
{Baker v. Selden, 101 US 99 (1880)}.

\bibitem{balganesh}
Shyamkrishna Balganesh.
\newblock The questionable origins of the copyright infringement analysis.
\newblock {\em Stanford Law Review}, 68:791--863, April 2016.

\bibitem{BalganeshMW}
Shyamkrishna Balganesh, Irina~D. Manta, and Tess Wilkinson-Ryan.
\newblock Judging similarity.
\newblock {\em Faculty Scholarship at Penn Law}, 1185, 2014.

\bibitem{verification_dilemmas}
Kenneth~A. Bamberger, Ran Canetti, Shafi Goldwasser, Rebecca Wexler, and
  Evan~J. Zimmerman.
\newblock Verification dilemmas, law, and the promise of zero-knowledge proofs.
\newblock {\em Berkeley Technology Law Journal}, 37, 2022.

\bibitem{Bellovin2014lawful}
Steven~M Bellovin, Matt Blaze, Sandy Clark, and Susan Landau.
\newblock Lawful hacking: Using existing vulnerabilities for wiretapping on the
  internet.
\newblock {\em Nw. J. Tech. \& Intell. Prop.}, 12:1, 2014.

\bibitem{BCTV14scalable}
Eli {Ben-Sasson}, Alessandro Chiesa, Eran Tromer, and Madars Virza.
\newblock Scalable zero knowledge via cycles of elliptic curves.
\newblock In {\em Proceedings of the 34th Annual International Cryptology
  Conference}, CRYPTO~'14, pages 276--294, 2014.

\bibitem{STOC:BCCT13}
Nir Bitansky, Ran Canetti, Alessandro Chiesa, and Eran Tromer.
\newblock Recursive composition and bootstrapping for {SNARKS} and
  proof-carrying data.
\newblock In Dan Boneh, Tim Roughgarden, and Joan Feigenbaum, editors, {\em
  45th ACM STOC}, pages 111--120. {ACM} Press, June 2013.

\bibitem{dukecaselaw}
James Boyle and Jennifer Jenkins.
\newblock {\em Intellectual Property: Law \& The Information Society---Cases
  and Materials}.
\newblock 5th edition, 2021.

\bibitem{butler1992pragmatism}
John~H Butler.
\newblock Pragmatism in software copyright: Computer associates v. altai.
\newblock {\em Harv. JL \& Tech.}, 6:183, 1992.

\bibitem{FORC:CohenDMS21}
Aloni Cohen, Moon Duchin, J.~N. Matthews, and Bhushan Suwal.
\newblock Census topdown: The impacts of differential privacy on redistricting.
\newblock In {\em {FORC}}, volume 192 of {\em LIPIcs}, pages 5:1--5:22. Schloss
  Dagstuhl - Leibniz-Zentrum f{\"{u}}r Informatik, 2021.

\bibitem{PNAS:CohenN20}
Aloni Cohen and Kobbi Nissim.
\newblock Towards formalizing the {GDPR}'s notion of singling out.
\newblock {\em Proc. Natl. Acad. Sci. {USA}}, 117(15):8344--8352, 2020.

\bibitem{cohen1986masking}
Amy~B Cohen.
\newblock Masking copyright decisionmaking: The meaninglessness of substantial
  similarity.
\newblock {\em UC Davis L. Rev.}, 20:719, 1986.

\bibitem{ca_v_altai}
{Computer Associates International, Inc. v. Altai, Inc., 982 F.2d 693 (1992)}.

\bibitem{concrete_v_classic}
{Concrete Machinery Co. v. Classic Lawn Ornaments, 843 F. 2d 600 (1st Cir.
  1988)}.

\bibitem{country_v_sheen}
{Country Kids ’N City Slicks, Inc. v. Sheen, 77 F.3d 1280 (10th Cir. 1996)}.

\bibitem{crowe1992}
Daniel~A Crowe.
\newblock Scope of copyright protection for non-literal design elements of
  computer software: Computer associates international, inc. v. altai, inc.
\newblock {\em . Louis ULJ}, 37:207, 1992.

\bibitem{davies_v_bowes}
{Davies v. Bowes, 209 F. 53 (S.D.N.Y. 1913)}.

\bibitem{eckes}
{Eckes v. Card Prices Update, 736 F.2d 859 (2d Cir. 1984)}.

\bibitem{effross1993assaying}
Walter~A Effross.
\newblock Assaying computer associates v. altai: How will the golden nugget
  test pan out.
\newblock {\em Rutgers Computer \& Tech. LJ}, 19:1, 1993.

\bibitem{mpc-book}
David Evans, Vladimir Kolesnikov, and Mike Rosulek.
\newblock A pragmatic introduction to secure multi-party computation.
\newblock {\em Foundations and Trends® in Privacy and Security},
  2(2-3):70--246, 2018.

\bibitem{SPW:FeigenbaumW18}
Joan Feigenbaum and Daniel~J. Weitzner.
\newblock On the incommensurability of laws and technical mechanisms: Or, what
  cryptography can't do.
\newblock In {\em Security Protocols Workshop}, volume 11286 of {\em Lecture
  Notes in Computer Science}, pages 266--279. Springer, 2018.

\bibitem{feist_v_rural}
{Feist Publications, Inc. v. Rural Telephone Service Co., 499 U.S. 340 (1991)}.

\bibitem{field}
Thomas~G. Field.
\newblock Fundamentals of intellectual property: Cases \& materials, 2012.

\bibitem{financial_v_moodys}
{Financial Inform. v. Moody's Investors Serv, 808 F.2d 204 (2d Cir. 1986)}.

\bibitem{fleming1969substantial}
Robert~Fuller Fleming.
\newblock Substantial similarity: Where plots really thicken.
\newblock In {\em Copyright L. Symp.}, volume~19, page 252. HeinOnline, 1969.

\bibitem{USENIX:FPSGW18}
Jonathan Frankle, Sunoo Park, Daniel Shaar, Shafi Goldwasser, and Daniel~J.
  Weitzner.
\newblock Practical accountability of secret processes.
\newblock In William Enck and Adrienne~Porter Felt, editors, {\em USENIX
  Security 2018}, pages 657--674. {USENIX} Association, August 2018.

\bibitem{EC:GarGolVas20}
Sanjam Garg, Shafi Goldwasser, and Prashant~Nalini Vasudevan.
\newblock Formalizing data deletion in the context of the right to be
  forgotten.
\newblock In Anne Canteaut and Yuval Ishai, editors, {\em EUROCRYPT~2020,
  Part~II}, volume 12106 of {\em {LNCS}}, pages 373--402. Springer, Heidelberg,
  May 2020.

\bibitem{gates_v_bando}
{Gates Rubber Co. v. Bando Chem. Indus., Ltd., 9 F.3d 823, 835 (10th Cir.
  1993)}.

\bibitem{WPES:GoldwasserP17}
Shafi Goldwasser and Sunoo Park.
\newblock Public accountability vs. secret laws: Can they coexist?: {A}
  cryptographic proposal.
\newblock In {\em WPES@CCS}, pages 99--110. {ACM}, 2017.

\bibitem{googacle}
{Google LLC v. Oracle America, Inc., 593 U.S. \_\_\_, 140 S. Ct. 520 (2021)}.

\bibitem{GRIGNETTI1964304}
Mario~C. Grignetti.
\newblock A note on the entropy of words in printed english.
\newblock {\em Information and Control}, 7(3):304--306, 1964.

\bibitem{ipbook}
James Grimmelmann.
\newblock Patterns of information law: Intellectual property done right, 2017.

\bibitem{EC:Groth16}
Jens Groth.
\newblock On the size of pairing-based non-interactive arguments.
\newblock In Marc Fischlin and Jean-S{\'{e}}bastien Coron, editors, {\em
  EUROCRYPT~2016, Part~II}, volume 9666 of {\em {LNCS}}, pages 305--326.
  Springer, Heidelberg, May 2016.

\bibitem{heer2004case}
Christopher Heer.
\newblock The case against copyright protection of non-literal elements of
  computer software.
\newblock {\em Intellectual Property Journal}, 18:1, 2004.

\bibitem{rosenthal_v_kalpakian}
{Herbert Rosenthal Jewelry Corp. v. Kalpakian, 446 F. 2d 738 (9th Cir. 1971)}.

\bibitem{boisson_v_banian}
{Judi Boisson and American Country Quilts and Linens, Inc. v. Banian Ltd., and
  Vijay Rao, 280 F.Supp.2d 10 (2003)}.

\bibitem{arXiv:JudFei22}
Samuel Judson and Joan Feigenbaum.
\newblock On heuristic models, assumptions, and parameters.
\newblock {\em CoRR}, abs/2201.07413, 2022.

\bibitem{arXiv:KKLNQTV21}
Seny Kamara, Mallory Knodel, Emma Llans{\'{o}}, Greg Nojeim, Lucy Qin, Dhanaraj
  Thakur, and Caitlin Vogus.
\newblock Outside looking in: Approaches to content moderation in end-to-end
  encrypted systems.
\newblock {\em CoRR}, abs/2202.04617, 2022.

\bibitem{SP:KMPQ21}
Seny Kamara, Tarik Moataz, Andrew Park, and Lucy Qin.
\newblock A decentralized and encrypted national gun registry.
\newblock In {\em 2021 {IEEE} Symposium on Security and Privacy}, pages
  1520--1537. {IEEE} Computer Society Press, May 2021.

\bibitem{architecture}
Jonathan~Seil Kim.
\newblock Filtering copyright infringement analysis in architectural works.
\newblock {\em University of Illinois Law Review}, page 281, 2018.

\bibitem{knitwaves_v_lollytogs}
{Knitwaves, Inc. v. Lollytogs Ltd.(Inc.), 71 F. 3d 996 (2nd Cir. 1995)}.

\bibitem{MOBISYS:KohNB21}
John~S. Koh, Jason Nieh, and Steven~M. Bellovin.
\newblock Encrypted cloud photo storage using google photos.
\newblock In {\em MobiSys}, pages 136--149. {ACM}, 2021.

\bibitem{kohus_v_mariol}
{Kohus v. Mariol, 328 F.3d 848 (6th Cir. 2003)}.

\bibitem{kolmogorov1963tables}
Andrei~N Kolmogorov.
\newblock On tables of random numbers.
\newblock {\em Sankhy{\=a}: The Indian Journal of Statistics, Series A}, pages
  369--376, 1963.

\bibitem{kregos_v_ap}
{Kregos v. Associated Press, 937 F.2d 700 (2d Cir. 1991)}.

\bibitem{USENIX:KulMay21}
Anunay Kulshrestha and Jonathan~R. Mayer.
\newblock Identifying harmful media in end-to-end encrypted communication:
  Efficient private membership computation.
\newblock In Michael Bailey and Rachel Greenstadt, editors, {\em USENIX
  Security 2021}, pages 893--910. {USENIX} Association, August 2021.

\bibitem{laroche}
Guillaume Laroche.
\newblock Striking similarities: Toward a quantitative measure of melodic
  copyright infringement.
\newblock {\em Int\'{e}gral: The Journal of Applied Musical Thought},
  25:39--88, 2011.

\bibitem{latman}
Alan Latman.
\newblock ``{P}robative similarity'' as proof of copying: Toward dispelling
  some myths in copyright infringement.
\newblock {\em Columbia Law Review}, 90, June 1990.

\bibitem{lemley}
Mark~A. Lemley.
\newblock Our bizarre system for proving copyright infringement.
\newblock {\em Journal of the Copyright Society}, 57, 2010.

\bibitem{DBLP:journals/iandc/Levin84}
Leonid~A. Levin.
\newblock Randomness conservation inequalities; information and independence in
  mathematical theories.
\newblock {\em Inf. Control.}, 61(1):15--37, 1984.

\bibitem{levin1973universal}
Leonid~Anatolevich Levin.
\newblock Universal sequential search problems.
\newblock {\em Problemy peredachi informatsii}, 9(3):115--116, 1973.

\bibitem{DBLP:books/daglib/0071317}
Ming Li and Paul M.~B. Vit{\'{a}}nyi.
\newblock {\em An introduction to Kolmogorov complexity and its applications}.
\newblock Texts and Monographs in Computer Science. Springer, 1993.

\bibitem{lim2021saving}
Daryl Lim.
\newblock Saving substantial similarity.
\newblock {\em Fla. L. Rev.}, 73:591, 2021.

\bibitem{lim}
Daryl Lim.
\newblock Saving substantial similarity.
\newblock {\em Florida Law Review}, 79, 2021.

\bibitem{lotus_v_borland}
{Lotus Dev. Corp. v. Borland Int'l, 49 F.3d 807 (1st Cir. 1995)}.

\bibitem{bender_v_west}
{Matthew Bender \& Co., Inc. v. West Pub. Co., 158 F.3d 693 (1998)}.

\bibitem{morrissey_v_pg}
{Morrissey v. Procter \& Gamble Company, 379 F. 2d 675 (1st Cir. 1967)}.

\bibitem{nichols_v_universal}
{Nichols v. Universal Pictures Corporation et al., 45 F.2d 119 (1930)}.

\bibitem{nimmer}
David Nimmer.
\newblock {\em Nimmer on copyright}.
\newblock Matthew Bender Elite Products, 2013.

\bibitem{nimmer1988structured}
David Nimmer, Richard~L Bernacchi, and Gary~N Frischling.
\newblock A structured approach to analyzing the substantial similarity of
  computer software in copyright infringement cases.
\newblock {\em Ariz. St. LJ}, 20:625, 1988.

\bibitem{PODS:Nissim21}
Kobbi Nissim.
\newblock Privacy: From database reconstruction to legal theorems.
\newblock In {\em {PODS}}, pages 33--41. {ACM}, 2021.

\bibitem{Nissim2017bridging}
Kobbi Nissim, Aaron Bembenek, Alexandra Wood, Mark Bun, Marco Gaboardi, Urs
  Gasser, David~R O'Brien, Thomas Steinke, and Salil Vadhan.
\newblock Bridging the gap between computer science and legal approaches to
  privacy.
\newblock {\em Harv. JL \& Tech.}, 31:687, 2017.

\bibitem{osterberg}
Robert~C Osterberg and Eric~C Osterberg.
\newblock {\em Substantial Similarity in Copyright Law}.
\newblock PLI Press, 2003.

\bibitem{ParchomovskyG}
Gideon Parchomovsky and Kevin~A. Goldman.
\newblock Fair use harbors.
\newblock {\em Faculty Scholarship at Penn Law}, 173, 2007.

\bibitem{peter_pan}
{Peter Pan Fabrics, Inc. v. Martin Weiner Corp., 274 F.2d 487, 489 (2d Cir.
  1960)}.

\bibitem{ready_v_cantrell}
{R. Ready Productions, Inc. v. Cantrell, 85 F. Supp. 2d 672 (Dist. Court, SD
  Texas 2000)}.

\bibitem{roodhuyzen}
Nicole~K. Roodhuyzen.
\newblock Do we even need a test? a reevaluation of assessing substantial
  similarity in a copyright infringement case.
\newblock {\em Journal of Law and Policy}, 15, 2007.

\bibitem{roth_v_united}
{Roth v. United States, 354 U.S. 476, 77 S. Ct. 1304 (1957)}.

\bibitem{samuelson}
Pamela Samuelson.
\newblock A fresh look at tests for nonliteral copyright infringement.
\newblock {\em Northwestern University Law Review}, 107:1821, 2013.

\bibitem{samuelson2016}
Pamela Samuelson.
\newblock Functionality and expression in computer programs: Refining the tests
  for software copyright infringement.
\newblock {\em Berkeley Tech. LJ}, 31:1215, 2016.

\bibitem{samuelson2016reconceptualizing}
Pamela Samuelson.
\newblock Reconceptualizing copyright's merger doctrine.
\newblock {\em J. Copyright Soc'y USA}, 63:417, 2016.

\bibitem{samuelson1994}
Pamela Samuelson, Randall Davis, Mitchell~D Kapor, and Jerome~H Reichman.
\newblock Manifesto concerning the legal protection of computer programs, a.
\newblock {\em ColUM. l. reV.}, 94:2308, 1994.

\bibitem{sayood2017introduction}
Khalid Sayood.
\newblock {\em Introduction to data compression}.
\newblock Morgan Kaufmann, 2017.

\bibitem{USENIX:SchVar21}
Sarah Scheffler and Mayank Varia.
\newblock Protecting cryptography against compelled self-incrimination.
\newblock In Michael Bailey and Rachel Greenstadt, editors, {\em USENIX
  Security 2021}, pages 591--608. {USENIX} Association, August 2021.

\bibitem{sega_v_accolade}
{Sega Enterprises Ltd. v. Accolade, Inc., 977 F 2d 1510 (9th Cir. 1992)}.

\bibitem{shannon1951prediction}
Claude~E Shannon.
\newblock Prediction and entropy of printed english.
\newblock {\em Bell system technical journal}, 30(1):50--64, 1951.

\bibitem{shaw_v_lindheim}
{Shaw v. Lindheim, 919 F. 2d 1353 (9th Cir. 1990)}.

\bibitem{sheldon_v_metro}
{Sheldon v. Metro-Goldwyn Pictures Corp., 81 F. 2d 49, 54 (CA2 1936)}.

\bibitem{krofft_v_mcdonalds}
{Sid \& Marty Krofft TV Prods. v. McDonald's Corp., 562 F.2d 1157 (9th Cir.
  1977)}.

\bibitem{smith2019truth}
Cathay~YN Smith.
\newblock Truth, lies, and copyright.
\newblock {\em Nev. Law Journal}, 20:201, 2019.

\bibitem{softei_v_dragon}
{Softei v. Dragon Medical \& Scientific Communications Inc., 118 F.3d 955 (Fed.
  Cir. 1997)}.

\bibitem{southco_v_kanebridge}
{Southco, Inc. v. Kanebridge Corp., 390 F.3d 276 (2004)}.

\bibitem{stanfield}
B.~MacPaul Stanfield.
\newblock Finding the fact of familiarity: Assessing judicial similarity tests
  in copyright infringement actions.
\newblock {\em Drake Law Review}, 49:489--512, 2001.

\bibitem{transwestern_v_multimedia}
{Transwestern Publ’g Co. v. Multimedia Mktg. Assocs., Inc., 133 F.3d 773
  (10th Cir. 1998)}.

\bibitem{fox_v_stonesifer}
{Twentieth Century-Fox Film Corp. v. Stonesifer, 140 F. 2d 579 (9th Circuit
  1944)}.

\bibitem{title17}
{U.S.\ Copyright Office}.
\newblock Copyright law of the {United States} ({Title} 17) and related laws
  contained in {Title} 17 of the {United States} code.
\newblock \url{https://www.copyright.gov/title17/}, May 2021.

\bibitem{Valiant08inc}
Paul Valiant.
\newblock Incrementally verifiable computation or proofs of knowledge imply
  time/space efficiency.
\newblock In {\em Proceedings of the 5th Theory of Cryptography Conference},
  TCC~'08, pages 1--18, 2008.

\end{thebibliography}
\bibliographystyle{plain}

\end{document}